\documentclass[twocolumn,showpacs,pre,floatfix]{revtex4}
\usepackage{amsmath,amsfonts,amssymb}
\usepackage{graphicx} 

\begin{document}
\title{Thermalization of an anisotropic granular particle}
\author{P. Viot$^1$}
\author{J. Talbot$^2$} 
\affiliation{$^1$Laboratoire de Physique
Th\'eorique des Liquides, Universit\'e Pierre et Marie Curie, 4, place
Jussieu, 75252 Paris Cedex, 05 France}
\affiliation{$^2$Department   of  Chemistry  and  Biochemistry,
 Duquesne University, Pittsburgh, PA 15282-1530}

\begin{abstract}
We investigate  the dynamics of  a  needle  in a two-dimensional  bath
composed   of thermalized   point  particles.   Collisions between the
needle  and   points  are inelastic  and   characterized  by  a normal
restitution coefficient $\alpha<1$. By using the Enskog-Boltzmann equation,
we obtain analytical expressions for  the translational and rotational
granular  temperatures of  the needle  and   show that  these  are, in
general,   different   from the   bath  temperature. The translational
temperature always exceeds the rotational  one, though the  difference
decreases with increasing moment  of inertia. The predictions of  the
theory are  in very good agreement  with numerical  simulations of the
model.
\end{abstract}
\maketitle
\section{Introduction}

The dissipative nature of the collisions  in granular systems leads to
fundamentally different behavior   from  their thermal   analogues.  A
striking example  is the   lack of  energy equipartition  between  the
degrees of freedom in granular systems. For example experiments
\cite{WP02,FM02} and computer simulations \cite{BT02,BT02a,KT03} have
shown that in binary   mixtures of isotropic inelastic  particles, the
granular temperatures of the two species are different.

These results have  prompted theoreticians to  investigate some simple
model systems.  For  example, Martin and  Piasecki\cite{MP99} examined
the behavior of a spherical  tracer particle immersed in a homogeneous
fluid  in  equilibrium  at a  temperature  $T$.  They showed that  the
Enskog-Boltzmann   equation  of   the   tracer particle  possesses   a
stationary Maxwellian  velocity    distribution  characterized  by  an
effective temperature that is smaller than $T$.

Although many granular  systems contain particles  that are manifestly
anisotropic, most studies have been confined to spherical particles. A
notable exception is the work of Aspelmeier, Huthmann and Zippelius
\cite{AHZ00,HAZ99} that examines the free cooling of an assembly of
inelastic needles with the aid of a pseudo-Liouville operator. They
predicted an exponentially fast cooling followed by a state with a
stationary ratio of translational and rotational energy.  This two
stage cooling was confirmed by event driven simulations.

In  this work we  examine the  breakdown of  equipartition in a steady
state granular system containing an anisotropic particle. Motivated by
the studies of  Martin and Piasecki\cite{MP99} and  Aspelmeier, Huthmann
and   Zippelius\cite{AHZ00,HAZ99},  we   consider an  infinitely  thin
inelastic needle immersed in a bath of point particles.  Starting from
the  collision  rules of this  model,  we  derive the Enskog-Boltzmann
equation.  By assuming that  the points are   thermalized and that the
velocity   and  angular  velocity distributions  of    the  needle are
Maxwellian, we derive analytical expressions for the translational and
rotational granular  temperatures as a  function of the  masses of the
two species,  the moment of  inertia   of the  needle  and the  normal
restitution coefficient. Both these temperatures are smaller than that
of the bath.  The  rotational  granular temperature is usually   lower
than  the translational one, except  for very light bath particles for
which the two are equal.

We also report essentially  exact numerical results obtained using  a
stochastic simulation  method. The  theory is  in  very good agreement
with the simulation, which  validates the assumption of the Maxwellian
shape of the distribution functions.

\section{Model}
\begin{figure}
\resizebox{8cm}{!}{\includegraphics{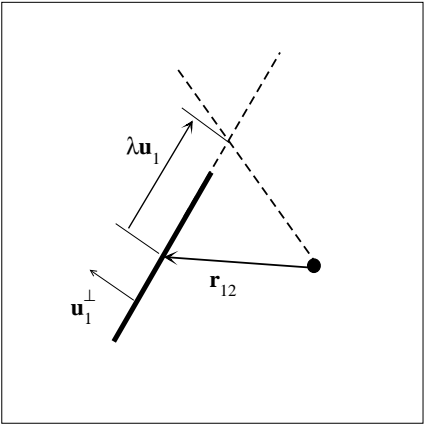}}

\caption{Geometry of the needle and a point in the plane: ${\bf
r}_{12}$ denotes  a  vector joining  the  point  labeled $2$ and  the
center of the needle;  ${\bf u}_1$ is a  unit vector along the axis of
the needle, $\lambda$ is  the algebraic distance  between the center  of the
needle  and the  point  of impact and ${\bf  u}_1^\perp$  is a unit vector
perpendicular to the axis of the needle. For a collision to occur one
requires that $|\lambda|\leq L/2$ when the point lies on the line defined by
the needle; i.e. when Eq.(\ref{eq:1}) is satisfied.}
\label{fig:1} 
\end{figure}

We examine a two-dimensional  system consisting of an  infinitely thin
needle of length $L$, mass $M$ and moment of inertia $I$ immersed in a
bath of point particles each of mass $m$.   The vector position of the
center of mass of the needle and a point particle are denoted by ${\bf
r}_1$ and ${\bf r}_2$, respectively. The  orientation of the needle is
specified by  a unit  vector ${\bf  u}_1$ that  points along its axis.
Let ${\bf  r}_{12}={\bf r}_1-{\bf r}_2$  and  ${\bf u}_{1}^\perp$ denote a
vector perpendicular to  ${\bf u}_1$. A  collision between a needle
and a point occurs when
\begin{equation}\label{eq:1}
{\bf r}_{12}.{\bf u}_{1}^\perp=0
\end{equation}
and  $|\lambda|<L/2$. (see
Fig.~\ref{fig:1}). The relative velocity of the point of
contact ${\bf V}$  is given by 
\begin{equation}\label{eq:2}
{\bf V}={\bf v}_{12}+\lambda\dot{{\bf u}}_1.
\end{equation}

The pre- and post-collisional quantities (the latter are labeled with
a prime) obey the usual conservation laws:
\begin{itemize}			
\item Total momentum conservation
\begin{equation}\label{eq:3}
M {\bf v'}_{1}+m{\bf v'}_2=M {\bf v}_{1}+m{\bf v}_2.
\end{equation}
\item Angular momentum conservation with respect to the point of
contact
\begin{equation}\label{eq:4}
I\omega'_1{\bf k}=I\omega_1{\bf k}+M\lambda{\bf u}_1\land\left({\bf v'}_{1}-{\bf v}_{1}\right),
\end{equation}
where ${\bf k}$ is a unit vector perpendicular to the plane such that
${\bf k}={\bf u}_1\land {\bf u}_1^\perp$. 
\end{itemize}
As a result of the collision, the  relative velocity of the contacting
points changes instantaneously according to the following relations:
\begin{align}\label{eq:5}
{\bf V'}.{\bf u}_{1}^\perp &=-\alpha{\bf V}.{\bf u}_{1}^\perp \\
{\bf V'}.{\bf u}_{1} &={\bf V}_.{\bf u}_{1} \label{eq:6}
\end{align}
where $\alpha$ denotes the normal restitution coefficient.  For the sake of
simplicity we have taken the  tangential restitution coefficient equal
to one. This choice is reflected in the form of Eq.~(\ref{eq:6}).

By  combining Eqs.~(\ref{eq:2})-~(\ref{eq:5}) one  obtains, after some
algebra, the change of the needle momentum  $\Delta {\bf p}=M ({\bf v}'_1-
{\bf v}_1)$
\begin{equation}\label{eq:7}
\Delta{\bf          p}.{\bf        u}_{1}^\perp=-\frac{(1+\alpha){\bf        V}.{\bf
u}_{1}^\perp}{\frac{1}{m}+\frac{1}{M}+\frac{\lambda^2}{I}}.
\end{equation}
The  inelastic collision leads  to  a  loss  of translational  kinetic
energy,
\begin{align}\label{eq:8}
 \Delta E_{1}^T&=\frac{M}{2}\left(({\bf v}'_{1})^2-({\bf v}_{1})^2\right)\nonumber\\
&=\Delta {\bf p}.{\bf v}_{1}+\frac{1}{M}\frac{\Delta {\bf p}^2}{2}\nonumber\\
&=-(1+\alpha)\frac{{\bf V}.{\bf u}_{1}^\perp{\bf v}_1.{\bf u}_{1}^\perp.}
{\frac{1}{m}+\frac{1}{M}+\frac{\lambda^2}{I}}+\frac{M}{2}
\frac{(1+\alpha)^2({\bf V}.{\bf u}_{1}^\perp)^2}
{\left(\frac{1}{m}+\frac{1}{M}+\frac{\lambda^2}{I}\right)^2},
\end{align}
and rotational energy,
\begin{align}
\Delta E_1^R&=\frac{I}{2}\left((\omega'_1)^2-(\omega_1)^2\right)\nonumber\\
&=-\frac{\lambda(1+\alpha)}{2}\frac{{\bf V}.{\bf u}_{1}^\perp(\omega_1'+\omega_1)}
{\frac{1}{m}+\frac{1}{M}+\frac{\lambda^2}{I}}\nonumber\\
&=-\lambda(1+\alpha)\frac{{\bf V}.{\bf u}_{1}^\perp\omega_1}
{\frac{1}{m}+\frac{1}{M}+\frac{\lambda^2}{I}}+
\frac{\lambda^2(1+\alpha)^2({\bf V}.{\bf u}_{1}^\perp)^2}
{I\left(\frac{1}{m}+\frac{1}{M}+\frac{\lambda^2}{I}\right)^2}.
\end{align}

\section{Pseudo-Liouville equation}
For particles  with  hard-core interactions, the  kinetic evolution of
$N$-particle distribution $f({\bf r}^N,{\bf  v}^N)$ is described  by a
pseudo-Liouville operator, where ${\bf  r}^N$ is a short-hand notation
for the positions (and internal  degrees of freedom, i.e.  angle $\theta_1$
with the $x$-axis for the needle) of $N$ particles and ${\bf v}^N$ for
their velocities (and angular velocities  for the needle).  Originally
derived for spheres and elastic collisions\cite{EDHV69}, the formalism
has been  extended to systems  composed  of inelastic particles. Since
collisions  are instantaneous\cite{VNE00,BMD96}, the  pseudo-Liouville
operator $L(    {\bf r}^N,{\bf  v}^N)=L^0({\bf r}^N)+\sum_{i<j}\overline{
T}_{ij}$ is the sum of the  free particle streaming operator $L^0({\bf
r}^N)$ and  of  the sum  of  two-body collision operators  $\overline{
T}_{ij}$. Since  we  are  interested  in  the homogeneous state,   the
distribution  function $f({\bf v}_1,\omega_1)$ of  the  needle system is
described by
\begin{equation}\label{eq:9}
\frac{\partial f( {\bf v}_1,\omega_1)}{\partial t}=N \int\frac{d\theta_1}{2\pi}\int d{\bf v_2}\int d{\bf r}_2
\overline{ T}_{12}f( {\bf v}_1,\omega_1,{\bf v}_2)
\end{equation} 
where $N$ is the total number of points, $f( {\bf v}_1,\omega_1,{\bf v}_2)$
the distribution function of the needle and a point, and $\overline{
T}_{12}$ is the collision operator between a needle and a point.

The assumption of  molecular chaos yields a factorization of the
two body distribution $f( {\bf v}_1,\omega_1,{\bf v}_2)=f( {\bf v}_1,\omega_1)
\Phi ({\bf v}_2)$, where $\Phi({\bf v}_2)$ denotes the point distribution function.
Since the point particles are thermalized, a stationary state for the
system (tracer needle and points) can be reached, and one finally
obtains for the needle distribution 
\begin{equation}\label{eq:10}
\int d\theta_1\int d{\bf v_2}\int d{\bf r}_2\overline{ T}_{12}f({\bf v}_1,{\bf \omega}_1) \Phi ({\bf v}_2)=0
\end{equation}
where 
\begin{equation}
\Phi ({\bf v}_2)\propto \exp\left(-\frac{m{\bf v}_2^2}{2T}\right)
\end{equation}
where $T$ is the temperature of the bath.

To build the point-needle collision operator, $\overline{ T}_{12}$, one
must include  the change   in  quantities (i.e. velocity   and angular
momentum)  produced during  the   infinitesimal time  interval of  the
collision.  This operator  is different  from   zero only  if the  two
particles are  in contact and if the  particles were  approaching just
before the collision\cite{AHZ00}. The explicit form of the operator is
\begin{align}\label{eq:11}
\overline{ T}_{12}&\propto \Theta(L/2-|\lambda|)\delta(|{\bf r}_{12}.{\bf u}_{1}^\perp|-0^+)\nonumber\\
&\left|\frac{d|{\bf r}_{12}.{\bf u}_{1}^\perp|}{dt}\right|
\Theta \left(-\left|\frac{d|{\bf r}_{12}.{\bf u}_{1}^\perp|}{dt}\right|\right)(b_{12}-1),
\end{align}
where $b_{12}$ is an operator  that changes pre-collisional quantities
to post-collisional quantities and $\Theta(x)$ is the Heaviside function.

The other terms of the  collision operator correspond to the necessary
conditions     of     contact   $ \Theta    (L/2-|\lambda|)\delta(|{\bf   r}_{12}.{\bf
r}_{1}^\perp|-0^+)$,      and     approach    $\Theta\left(-\left|\frac{d|{\bf
r}_{12}.{\bf u}_{1}^\perp|}{dt}\right|\right)$.

For an  isotropic tracer particle in  an isotropic particle  bath, Martin
and  Piasecki\cite{MP99}  solved    the  stationary   Enskog-Boltzmann
equation,   showing  that the    velocity  distribution of  the tracer
particle remains gaussian in  a thermalized bath.   It is worth noting
that for finite dilutions, the  velocity distribution function is  not
Maxwellian\cite{BT02},  but the deviations  from the gaussian shape of
the distribution function of the velocities (that can be calculated in
a perturbative way)  yields small corrections  to the estimate of  the
granular temperature.  More significant  deviations  are expected  for
finite dilution systems\cite{BMP02}.

For  a mixture of a  needle and  points, one  can  show that a Maxwell
distribution of the needle angular and translational momenta cannot be
a solution  of the stationary Enskog-Boltzmann  equation even  for the
case of an infinitely diluted needle. This is due to the fact that the
change of momentum depends  on the location of the  point of impact on
the  needle   (the right-hand  side    of Eq.~(\ref{eq:7}) depends  on
$\lambda$). However, we  impose gaussian distributions for the translational
and  angular   velocities of the   needle. We show below that this is
a very good approximation.

In the   stationary state, the  loss  of translational  and rotational
kinetic  energies is on average zero and can be expressed by means of
the collision operator
\begin{align}
\langle \Delta  E_1^T\rangle&=\langle  \overline{ T}_{12}  f({\bf v}_1,{\bf \omega}_1) \Phi({\bf v}_2)    E_1^T({\bf
v}_1)\rangle =0\nonumber\\
\langle \Delta  E_1^R\rangle&=\langle   \overline{ T}_{12}  f({\bf v}_1,{\bf \omega}_1) \Phi({\bf
v}_2)  E_1^R({\bf \omega}_1) \rangle =0
\end{align}
where the brackets denote an average over the independent variables.
This corresponds to taking the second moments of  the distribution function of
the  needle.   After     substitution  of   the collision     operator
(Eq.(\ref{eq:11})),  one obtains explicitly for  the translational kinetic energy

\begin{align}\label{eq:12}
&\int...\int  d{\bf r}_2 d{\bf v}_1 d{\bf v}_2 d\omega_1 d\theta_1  \Theta(L/2-|\lambda|)\delta(|{\bf r}_{12}.{\bf u}_{1}^\perp|-0^+)\nonumber\\
&\left|\frac{d|{\bf r}_{12}.{\bf u}_{1}^\perp|}{dt}\right|
\Theta\left(-\left|\frac{d|{\bf r}_{12}.{\bf u}_{1}^\perp|}{dt}\right|\right) f( {\bf v}_1,{\bf \omega}_1)\Phi({\bf v}_2)  \Delta E^T_1 =0.
\end{align}

A similar   equation can be written for   rotational kinetic energy. The
needle   distribution function $f({\bf v}_1, \omega_1)$ is
then given by 
\begin{equation}\label{eq:13}
f({\bf v}_1, \omega_1)\propto \exp\left(-\frac{M{\bf v_1}^2}{2\gamma_TT}-\frac{I{\bf \omega_1}^2}{2\gamma_RT}\right),
\end{equation}
where  $\gamma_T$  and   $\gamma_R$ are  the  ratios of   the  translational and
rotational needle temperatures to the bath temperature, respectively.

By inserting Eq.~(\ref{eq:13}) in Eq.~(\ref{eq:12}) and in the
corresponding equation for the rotational energy one obtains, after
some tedious but straightforward calculation (outlined in Appendix A), the following set of
equations
\begin{align}\label{eq:14}
\int_0^1dx b\frac{\sqrt{1+akx^2}}{1+kx^2}&=\frac{1+\alpha}{2}\int_0^1dx
\frac{(1+akx^2)^{3/2}}{(1+kx^2)^2},\\\label{eq:15}
\int_0^1dx ax^2\frac{\sqrt{1+akx^2}}{1+kx^2}&=\frac{1+\alpha}{2}\int_0^1dx
\frac{x^2(1+akx^2)^{3/2}}{(1+kx^2)^2},
\end{align}
where 
\begin{equation}
k=\frac{L^2}{4I\left(\frac{1}{m}+\frac{1}{M}\right)},
\end{equation}
\begin{equation}\label{eq:16}
a=\gamma_R\frac{(M+m)}{M+m\gamma_T},
\end{equation}
and 
\begin{equation}\label{eq:17}
b=\gamma_T\frac{M+m}{M+m\gamma_T}.
\end{equation}

Explicit  expressions    for the integrals  appearing  in
Eqs.~(\ref{eq:14})    and    (\ref{eq:15})  are    given   in  Appendix
B. Eq.~(\ref{eq:15}) is an implicit equation for $a$ that, for a given
value of $\alpha$, can  be solved with  standard numerical methods.  $b$ is
then  easily   obtained  by calculating   the   ratio of integrals  of
Eq.~(\ref{eq:14}). Finally,  from the values of $a$  and $b$, $\gamma_T$ and
$\gamma_R$ can be obtained from Eqs.(\ref{eq:16})-(\ref{eq:17}).

\section{Results}
We first note that for  the elastic case, $\alpha=1$,  we have the solution
$a=b=1$ which  gives $\gamma_T=1$  and  since $a/b=\gamma_R/\gamma_T$,  $\gamma_R=1$ which
corresponds to the equilibrium case\cite{FD81,FD83}.

\begin{figure}
\resizebox{8cm}{!}{\includegraphics{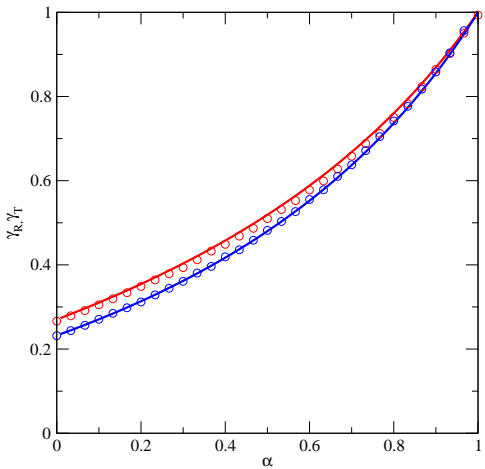}}

\caption{Ratio of the translational (full curve) $\gamma_T$ and
rotational  $\gamma_R$     (dashed  curve)  granular    temperature to  the
temperature of the bath  versus the normal restitution coefficient $\alpha$
for a homogeneous needle with $M=m$.}\label{fig:2}
\end{figure}

Figure \ref{fig:2} shows the ratio of the translational temperature to
the bath temperature  and  of the rotational  temperature  to the bath
temperature for a homogeneous needle ($I=ML^2/12$) whose mass is equal
to the mass of a bath particle. As  the normal restitution coefficient
decreases from $1$ to $0$ both  ratios decrease monotonically from $1$
to a strictly positive value, such  that the translational temperature
is always larger the rotational  temperature.  An analogous  situation
has   been  noted  for    the  free  cooling  of  needles   in   three
dimensions\cite{AHZ00}.

\begin{figure} \resizebox{8cm}{!}{\includegraphics{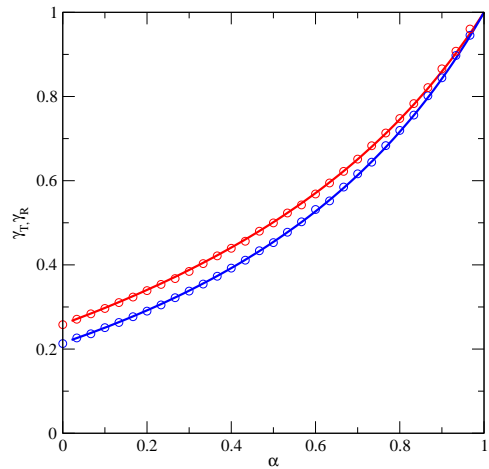}}
\caption{Ratio of the translational (full curve) $\gamma_T$ and
rotational $\gamma_R$ (dashed curve) granular temperature to the
temperature of the bath versus the normal restitution coefficient
$\alpha$ for an inhomogeneous needle ($I=ML^2/16$)with
$M=m$.}\label{fig:3} \end{figure}

Figure \ref{fig:3} shows  the same ratios  for an inhomogeneous needle
($I=ML^2/16$) corresponding to  relatively lighter ends and a heavier
center. Again the translational  temperature is always larger than the
rotational temperature, and the   differences are greater than  for  a
homogeneous needle.    The  difference  between  the two  temperatures
vanishes in  the limit of very  large moment of  inertia. Contrary to
the   free  cooling state   of   needles  \cite{AHZ00},  there is   no
``critical'' value of the moment of inertia above which the rotational
temperature is larger than the translational temperature.

We now return to consideration of a homogeneous needle and investigate
the effect   of varying  the  mass  ratio, $m/M$,  at  constant normal
restitution  coefficient.     When   $m/M\to0$,    $k\to0$,   which  gives
$a=b=(1+\alpha)/2$.  Hence,  for very  light bath particles,  equipartition
between  translational   and    rotational  granular  temperatures  is
asymptotically obtained and we have
\begin{equation}\label{eq:18}
\gamma_R=\gamma_T=\frac{1+\alpha}{2}.
\end{equation}
This same limit is obtained for a spherical tracer particle
\cite{MP99}. We conjecture that the result is independent of the 
shape of the  tracer  particle and the dimension. 

\begin{figure} \resizebox{8cm}{!}{\includegraphics{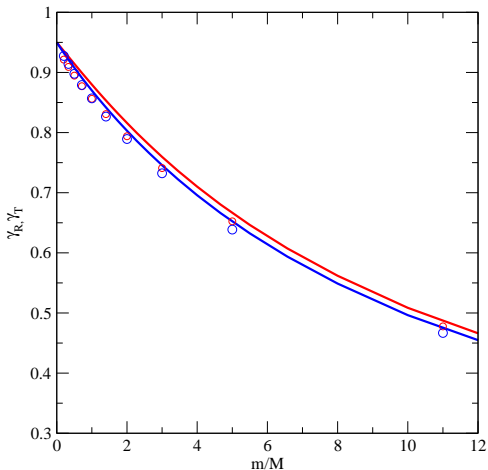}}
\caption{Ratio of the translational (full curve) $\gamma_T$ and
rotational $\gamma_R$ (dashed curve) granular temperature to the
temperature of the bath versus the ratio of the masses $m/M$ for a
homogeneous needle  ($I=ML^2/12$) and $\alpha=0.9$.}\label{fig:4}
\end{figure}

Figure  \ref{fig:4} displays the variation with  the mass ratio of the
translational and  rotational temperature of   the tracer needle. Note
that when the mass of the bath particles is very small compared to the
needle  particle, the    two   curves go     to the  limit    given by
Eq.~(\ref{eq:18}).

\section{Simulation} 

In  order to assess the accuracy  of the theory  we developed a Direct
Simulation  Monte Carlo (DSMC) code  to obtain  numerical solutions of
the    non-linear  homogeneous Boltzmann   equation.   The  simulation
generates a sequence  of collisions   with  a variable  time  interval
between each.

At  the beginning of  each step, we first calculate  the total flux of
colliding particles on both sides of the needle which is a function of
the component of the   velocity of the center of   mass normal to  the
length   of the  needle   ${\bf v}_1.{\bf u}_{1}^\perp$  and the  angular
velocity $\omega$. Next a  waiting time  to  the next collision is  sampled
from an exponential distribution with rate constant equal to the total
flux. The  needle is  rotated at  its current  angular velocity to the
point of  collision. The  location  of the  collision is then selected
with a  probability proportional to the  position dependent  flux. The
ratio  of the flux  on the left  hand side  to the total  flux is then
compared to a uniform random number between zero and one. If the ratio
is greater than the random number, the collision  occurs on the right;
otherwise is occurs on the left. The normal component of the colliding
point  is then selected from a   gaussian flux. Finally, the collision
rules are applied and the normal  velocity and angular velocity of the
needle are updated.

The simulation results for $10^7$ collisions per run are compared with
the theory  in Figs.~\ref{fig:2}-~\ref{fig:3}. Although the  theory is
not  exact,  the agreement  is   excellent.  The  discrepancy  between
simulation and theory is never greater than $1\%$ whatever the value of
the restitution coefficient. In Fig.~\ref{fig:4}, when the mass of the
needle increases, the granular temperatures obtained by simulation are
 slightly smaller than the analytical result.

\begin{figure}
\resizebox{8cm}{!}{\includegraphics{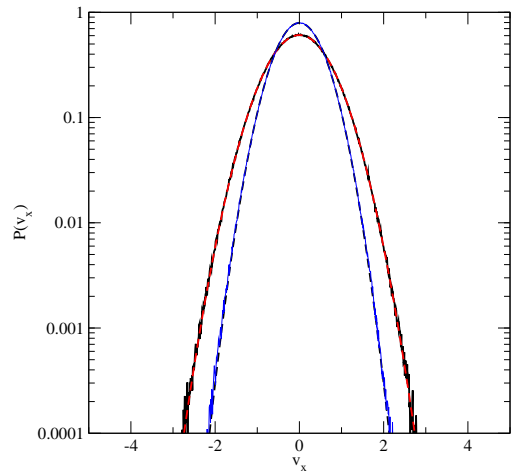}}

\caption{Distribution of the normal velocity component of the 
needle  for two values  of the normal restitution coefficient, $\alpha=0.5$
and $\alpha=0.9$ (broader curve).  The solid (noisy)  curves are the simulation results and
the   dashed curves correspond to  Gaussian  distributions of velocity
evaluated for the corresponding theoretical granular temperature.}\label{fig:5}
\end{figure}

\begin{figure}
\resizebox{8cm}{!}{\includegraphics{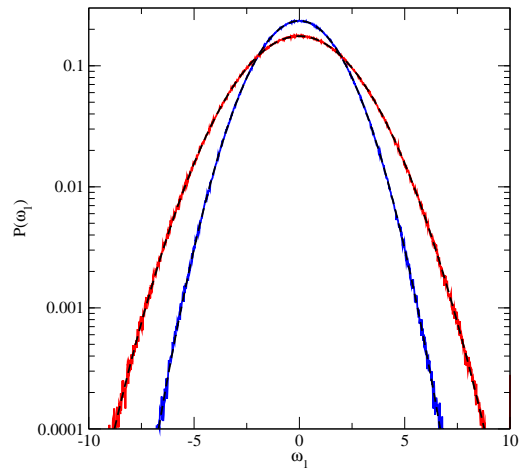}}

 \caption{Same as    \ref{fig:5}, except  the   angular velocity    is
 shown.}\label{fig:6}
\end{figure}

Fig.~\ref{fig:5} and   Fig.~\ref{fig:6}  show  the   translational and
angular velocity distributions, respectively, for two different values
of the normal restitution coefficient ($0.5$ and  $0.9$). On the scale
of the plot, the simulated distributions are in quantitative agreement
with gaussian  distributions  at the corresponding  theoretical  granular
temperature,  confirming the validity of employing   the latter in the
theory.


\section{Conclusion} 
We have   shown that    the   translational and  rotational   granular
temperatures  of an anisotropic tracer particle  immersed in a bath of
point particles depend on the ratio  of the masses of  a point and the
needle and the moment of inertia. They are,  in general, not equal and
differ from   the bath   temperature.  In   addition, we  found   that
equipartition is   obtained asymptotically,  regardless of  the normal
restitution  coefficient, for very  light  bath particles.  We  expect
this   to be  a  general feature,  i.e.  in  higher dimensions and for
arbitrary shaped anisotropic particles.  It should be possible to test
this prediction experimentally.

\appendix \section{Needle average energy loss} As for binary mixtures
of spheres\cite{BT02}, we use  a gaussian ansatz for  the distribution
functions and  introduce two  different temperatures corresponding  to
the  translational and  rotational degrees  of  freedom of the needle.
The homogeneous distribution functions of the needle and of the points
are then  given respectively by 
\begin{equation} f({\bf v}_1,\omega_1)\sim
\exp\left(-\frac{M{\bf v}_1^2\gamma_T^{-1}}{2T}-\frac{I\omega_1^2\gamma_R^{-1}}{2T}\right),
\end{equation}

\begin{equation} \Phi({\bf v}_2)\sim
\exp\left(-\frac{m{\bf v}_2^2}{2T}\right),
 \end{equation}
where $T$ is the temperature of the bath, $\gamma_T$ and $\gamma_R$
the ratio of the translational (and rotational) temperature of the
needle to the bath temperature.

We introduce the vectors ${\boldsymbol \chi}$ and ${\boldsymbol \nu}$ such that
\begin{align}
{\boldsymbol \chi}&=\frac{1}{\sqrt{2T(M\gamma_T+m)}}\left(M{\bf v}_1+m{\bf v}_2\right)\\
{\boldsymbol \nu}&=\sqrt{\frac{mM}{2T(M\gamma_T+m)\gamma_T}}\left({\bf v}_1+\gamma_T{\bf v}_2\right)
\end{align}
The scalar product ${\bf V}.{\bf u}_1^\perp $ can be expressed as
\begin{align}\label{eq:19}
{\bf V}.{\bf u}_1^\perp&=\sqrt{\frac{2T}{mM\gamma_T}}\left[\left(\gamma_T-1\right)
{\boldsymbol \chi}.{\bf u}_1^\perp\right.\nonumber\\
&\left.+\sqrt{\gamma_T}\left(\sqrt{\frac{m}{M}}
+\sqrt{\frac{M}{m}}\right)
{\boldsymbol \nu}.{\bf u}_1^\perp\right]
\end{align}
Let us introduce $\xi=\omega_1\sqrt{\frac{I}{2T\gamma_R}}$.
The translational energy loss is given by the formula
\begin{align}\label{eq:20}
&\sum_{p=\pm1}\int d\lambda\int d\theta_1\int d{\boldsymbol \chi}\int d{\boldsymbol \nu}
\int d\omega_1\nonumber\\&\exp\left(-{\boldsymbol \chi}^2-{\boldsymbol \nu}^2-\xi^{2}\right)
|{\bf V}.{\bf u}_1^\perp|\Theta (p {\bf V}.{\bf u}_1^\perp) \Theta \left( |\lambda|-\frac{L}{2}\right) \Delta E_{1}^T.
\end{align}
Since Eq.~(\ref{eq:19})  depends only on ${\boldsymbol \chi}.{\bf u}_1^\perp$
and
${\boldsymbol \nu}.{\bf u}_1^\perp$ , one can freely integrate over the
direction of ${\bf u}_1$ for the vectors ${\boldsymbol \chi}$ and ${\boldsymbol \nu}$. The
integration over $\theta_1$ can be easily performed. If we introduce the
three dimensional vectors ${\bf G}$ and ${\bf s}$ with components:
\begin{align}
{\bf G}=&\left(G_1,G_2,G_3\right)\nonumber\\
=&\left(\sqrt{\frac{2T}{M\gamma_T+m}}(\gamma_T-1),\sqrt{\frac{2T\gamma_T}{M\gamma_T+m}}\frac{m+M}{\sqrt{mM}},\right.\nonumber\\
&\left.\lambda\sqrt{\frac{2T\gamma_R}{I}}\right)
\end{align}
and 
\begin{align}
{\bf s}=&\left(s_1,s_2,s_3\right)\nonumber\\
&\left({\boldsymbol \chi}.{\bf u}_1^\perp,{\boldsymbol \nu}.{\bf u}_1^\perp,\xi \right)
\end{align}
By inserting Eq.~(\ref{eq:8}) in Eq.~(\ref{eq:20}), the  average energy
loss  can be rewritten as
\begin{align}\label{eq:21}
&\sum_{p=\pm1}\int d\lambda\int d{\bf s}
\exp\left(-{\bf s}^2\right)
|{\bf G}.{\bf s}| \Theta (p{\bf G}.{\bf s})\Theta \left(|\lambda|-\frac{L}{2}\right)\nonumber\\&
\left[-\frac{(1+\alpha){\bf G}.{\bf s}}{\frac{1}{m}+\frac{1}{M}+
\frac{\lambda^2}{I}}\sqrt{\frac{2T}{M\gamma_T+m}}\left(\gamma_Ts_1+\sqrt{\frac{m\gamma_T}{M}}s_2\right)\right.\nonumber\\
&\left.+\frac{(1+\alpha)^2({\bf G}.{\bf s})^2}{2M\left(\frac{1}{m}+\frac{1}{M}+\frac{\lambda^2}{I}\right)^2}\right]
\end{align}
By defining a new coordinate system in which  the $z$-axis is parallel
to ${\bf G}$, one find that the 
integrals of Eq.~(\ref{eq:21}) involve  gaussian integrals of the form
\begin{align}
\int d{\bf s}\exp\left(-{\bf s}^2\right)
(|{\bf G}|s_z)^2 \Theta (\pm s_z)G_is_z=\frac{\pi}{8}|{\bf G}|^2G_i
\end{align}
and 
\begin{align}
\int d{\bf s}\exp\left(-{\bf s}^2\right)(|{\bf G}|s_z)^3
 \Theta (\pm s_z)=\frac{\pi}{8}|{\bf G^3}|
\end{align}
which finally leads to  Eq.~(\ref{eq:14}). The equation for  rotational
energy is derived following exactly the same procedure.

\section{Integrals}

The integrals appearing in Eqs.~(\ref{eq:14}) and (\ref{eq:15}) can be
evaluated explicitly. For completeness, we give below their analytical
expressions
\begin{align}
I_1&=\int_0^1dx\frac{\sqrt{1+akx^2}}{1+kx^2}\nonumber\\
&=\sqrt{\frac{a}{k}}\ln(\sqrt{ak}+\sqrt{1+ak})\nonumber\\&+\sqrt{\frac{1-a}{k}}\arctan\left(\sqrt{\frac{(1-a)k}{1+ak}}\right)
\end{align}

The integral $I_2$ is defined as
\begin{equation}
 I_2=\int_0^1dx\frac{(1+akx^2)^{3/2}}{(1+kx^2)^2}
\end{equation}
and satisfies the equation
\begin{align}
 I_2=I_1-(a-1)k\left.\frac{\partial \int_0^1dx\frac{\sqrt{1+akx^2}}{1+dx^2}}{\partial
 d}\right|_{d=k}
\end{align}
which gives
 \begin{align}
I_2&=\frac{a^{3/2}}{\sqrt{k}}\ln(\sqrt{ak}+\sqrt{1+ak})+\frac{(1-a)\sqrt{1+ak}}{2(1+k)}\nonumber\\
&\frac{1+a-2a^2}{2\sqrt{k(1-a)}}\arctan\left(\sqrt{\frac{(1-a)k}{1+ak}}\right)
\end{align}
\begin{align}
I_3&=\int_0^1dx\frac{x^2\sqrt{1+akx^2}}{1+kx^2}\nonumber\\&=\frac{1}{k}\left(\int_0^1dx\sqrt{1+akx^2}-I_1\right)\\
&=\frac{\sqrt{ak+1}}{2k}\frac{1-2a}{\sqrt{a}k^{3/2}}+\ln(\sqrt{ak}+\sqrt{1+ak})\nonumber\\&-\frac{\sqrt{1-a}}{k^{3/2}}\arctan\left(\sqrt{\frac{(1-a)k}{1+ak}}\right)
\end{align}

Finally, one has
\begin{equation}
I_4=\int_0^1dx\frac{x^2(1+akx^2)^{3/2}}{(1+kx^2)^2}
\end{equation}
which satisfy the algebraic equation
\begin{align}
I_2+kI_4&=aJ_1+(a-1)I_1
\end{align}
which gives
\begin{align}
I_4&=
\frac{\sqrt{1+ak}(ak-1+2a)}{2k(k+1)}\nonumber\\
&+\frac{\sqrt{a}(3-4a)}{2k^{3/2}}
\ln(\sqrt{ak}+\sqrt{1+ak})\nonumber\\
&+\frac{\sqrt{1-a}(1-4a)}{2k^{3/2}}\arctan\left(\sqrt{\frac{(1-a)k}{1+ak}}\right)
\end{align}
%\bibliography{cooling}

\begin{thebibliography}{10}

\bibitem{WP02}
R.~D. Wildman and D.~J. Parker, Phys. Rev. Lett. {\bf 88},  064301  (2002).

\bibitem{FM02}
K. Feitosa and N. Menon, Phys. Rev. Lett. {\bf 88},  198301  (2002).

\bibitem{BT02}
A. Barrat and E. Trizac, Granul. Matter {\bf 4},  57  (2002).

\bibitem{BT02a}
A. Barrat and E. Trizac, Phys. Rev. E {\bf 66},  051303  (2002).

\bibitem{KT03}
P.~E. Krouskop and J. Talbot, cond-mat/0303263.

\bibitem{MP99}
P.~A. Martin and J. Piasecki, Europhys. Lett {\bf 46},  613  (1999).

\bibitem{AHZ00}
T. Aspelmeier, T.~M. Huthmann, and A. Zippelius,  in {\em Granular Gases},
  edited by S. Luding and T. Poschel (Springer, Berlin, 2000), Chap.~Free
  cooling of particles with rotational degrees of freedom, p.\ 31.

\bibitem{HAZ99}
M. Huthmann, T. Aspelmeier, and A. Zippelius, Phys. Rev. E {\bf 60},  654
  (1999).

\bibitem{EDHV69}
M.~H. Ernst, J.~R. Dorfman, W.~R. Hoegy, and J.~M.~J. van Leeuwen, Physica {\bf
  45},  127  (1969).

\bibitem{VNE00}
T. van Noije and M. Ernst,  in {\em Granular Gases}, edited by S. Luding and T.
  Poschel (Springer, Berlin, 2001), Chap.~Kinetic Theory of Granular Gases, p.\
  3.

\bibitem{BMD96}
J.J. Brey, F. Moreno, and J.W. Dufty, Phys. Rev. E {\bf 54},  445  (1996).

\bibitem{BMP02}
T. Biben, P.~A. Martin, and J. Piaseski, Physica A {\bf 310},  308  (2002).

\bibitem{FD81}
D. Frenkel and J.F. Maguire, Phys. Rev. Lett. {\bf 47},  1025  (1981).

\bibitem{FD83}
D. Frenkel and J. Maguire, Mol. Phys. {\bf 49},  503  (1981).

\end{thebibliography}

\end{document}